\documentclass[12pt]{paper}

\begin{document}

\title{Quantum mechanics as a macrorealistic theory}

\author{N L Chuprikov}

\maketitle
\begin{abstract}

As contrasted with physicists to idolize Bell's theorem and quantum nonlocality, we
argue that quantum mechanics (QM), in reality, respects the principles of a
macroscopic realism (PMRs). The current QM to tell us that "\ldots the state of a
system can be instantaneously changed by a distant measurement\ldots" cannot be
treated as a physical theory. Its key statements - that {\it the EPR-Bell experiments
to violate Bell's inequality verify nonlocality}, and {\it nonlocal correlations
respect special relativity} - are false. Both the EPR-Bell experiments and theorems to
support the "non-signalling principle" are based on the implicit assumption that all
quantum postulates and, in particular, Born's averaging rule are fully applicable to
Cat states. However, this is not the case. Introducing observables (e.g.,
correlations) for Cat states violates the correspondence principle. Pure (macro- and
micro-)Cat states must be governed both by the PMRs and superposition principle. Our
(macrorealistic) model of a one-dimensional completed scattering shows how these
principles coexist with each other, in the case of a one-electron micro-Cat state.

\end{abstract}

\newcommand{\ppp}{\mbox{\hspace{5mm}}}
\newcommand{\ooo}{\mbox{\hspace{3mm}}}
\newcommand{\ooa}{\mbox{\hspace{1mm}}}

\section{Introduction} \label{intro}

At present, due to the EPR-Bell and Schr\"odinger's cat (or, simply, Cat) paradoxes to
appear for Cat states - coherent superpositions of macroscopically distinct states
(CSMDSs) - there is a widespread viewpoint that the linear formalism of quantum
mechanics (QM) is inapplicable to the macro-world, that the superposition principle
contradicts Leggett's principles of a macroscopic realism (PMRs) \cite{Le1}.

However, this is not the case. In this paper we show that the Cat paradox results from
the inconsistency of the current description of Cat states with the correspondence
principle (CP). {\it The Cat paradox is a correspondence problem}, rather than the
measurement or macro-objectification one. This problem is shown to be surmountable
within the linear formalism of standard QM.

A macrorealistic model of a one-dimensional (1D) completed scattering (see
\cite{Ch1,Ch2}) suggests that, in order to obey the CP, Cat states must be considered
in QM as a particular class of pure states to represent an intermediate link between
usual pure states and statistical mixtures. By our terminology, they are {\it
combined} states. Both the superposition principle and the PMRs govern such states,
coexisting with each other: a macroscopic distinction between states to enter a CSMDS
implies that there should be experiments to allow measuring observables for each of
them, without destroying interference whose observation implies another experimental
scheme.

We begin our analysis with the current vision of the paradoxes, which is based
eventually on the {\it orthodox} interpretation (OI) of the wave-particle duality. Its
main statements are divided here into three lessons to complement each other.

\section{Three lessons of the orthodox interpretation of quantum mechanics} \label{lessons}

Of course, the main mystery of modern QM is 'nonlocality' and 'observer-dependence' of
Cat states. These features are usually associated with compound systems. However, the
notion of Cat states is applicable to a single electron, as well. Moreover, in fact,
the OI's attitude toward the {\it one-particle} wave function lies in the bottom of
the mystery of Cat states.

\subsubsection*{Lesson 1: A 1D completed scattering.}

By the first lesson, the one-electron wave function describes an electron in a single
experiment. An electron cannot be treated as corpuscle or wave. It is {\it both}
corpuscle {\it and} wave simultaneously; or it is {\it neither} corpuscle {\it nor}
wave (such an electron will be called here "the OI's electron").

The OI's electron scattered on a 1D potential barrier is literally in the
superposition of the states 'transmitted electron' and 'reflected electron'. Though
these two states occupy macroscopically distinct spatial regions, to say that a
scattered OI's electron is {\it either} transmitted {\it or} reflected by the barrier
is erroneous in principle. In fact one cannot conceive the OI's electron as a part of
the universe (= external physical world). For example (see, e.g., \cite{Mu0, Win} and
references therein), the time spent by the OI's electron in the barrier region has
physical sense only in the context of a particular experimental situation
(observer-dependence) and it can be anomalously short or even negative by value
(nonlocality).

\subsubsection*{Lesson 2: The EPR-Bell paradox.}

By the second lesson, Bell's assumption about the existence of local hidden variables
leads to the inequality which is violated by QM and experiment. So that Bell's
assumption is wrong and hence two electrons to be in the Cat state have no
pre-existing properties and exhibit nonlocal features: inspecting the state of one of
these electrons instantly changes that of its partner to be well far from the former.
It is proved (see, e.g., \cite{Gh2} and references therein) that this occurs without
sending faster than light signals - nonlocal correlations respect the "no-signaling
principle".

\subsubsection*{Lesson 3: The Schr\"odinger's cat paradox.}

The third lesson teaches us that the 'observer-dependence' and 'nonlocality' of the
OI's universe pass inevitably from its micro- to macro-level and hence the cat to be
in the Cat state is both alive and died simultaneously. This means that the
superposition principle contradicts the PMRs, because in our perception the cat always
appears in a definite state, i.e., it is {\it either} alive {\it or} died. This also
means that the Cat state is needed in a macro-objectification and the Cat paradox
should be considered as a macro-objectification problem.

By this lesson, QM must be replaced by another theory which would imply the existence
of some physical process to suppress, at the macro-level, the action of the
superposition principle and thereby to convert a pure Cat state into statistical
mixture.

\subsubsection*{The grand total of the lessons.}
Neither the superposition principle nor the PMRs govern the whole OI's universe; it is
not a (local observer-independent) physical world and hence it is needed in a
macro-objectification. So that there are neither first-principles to govern the
external physical world nor the world itself (see also \cite{Har}).

\section{Reconciling standard QM with the PMRs} \label{criticism}

Our next step is to show that, in reality, the superposition principle and PMRs govern
together Cat states, respecting each other. To show this, we have to reconsider the
above lessons in the reverse order.

\subsection{The preparation problem for OI's Cat states} \label{cat}

Let us consider the Schr\"odinger's thought experiment with an electron scattered on a
1D potential barrier (instead of a decaying nucleus), the vial with a poison and the
long-suffering cat. All are in a closed box. The relationship between them is assumed
to be causal. By Schr\"odinger, in this thought experiment the cat symbolizes the
pointer of a macroscopic measuring device and the vial with a poison is the symbol of
an amplifier - intermediate link between the electron and cat. The electron's source
and the vial are supposed to be far enough from the potential barrier, the vial being
in the transmission spatial region. That is, an electron takes part here in a 1D {\it
completed} scattering.

According to the usual practice of setting this thought experiment as a
quantum-mechanical problem, we shall consider the electron and cat as parts of the
compound system 'electron+cat' and suppose that this system is in a pure state
expressed in terms of the electron's and cat's states.

Let $|\psi_{tr}^{end}\rangle$ and $|\psi_{ref}^{end}\rangle$ be pure states of
transmitted and reflected electrons, respectively, providing that $\langle
\psi_{tr}^{end}|\psi_{tr}^{end}\rangle=\langle
\psi_{ref}^{end}|\psi_{ref}^{end}\rangle=1/2.$ Besides, let $|0\rangle_c$ and
$|1\rangle_c$ be normalized pure states of a died and alive cat, respectively. Then
the Cat state $|\Psi\rangle_{e+c}$ of the 'electron+cat' system be
\begin{eqnarray} \label{1}
|\Psi\rangle_{e+c}=|\psi_{tr}^{end}\rangle\cdot|0\rangle_c
+|\psi_{ref}^{end}\rangle\cdot|1\rangle_c.
\end{eqnarray}

As was said above, by the current vision of Schr\"odinger's thought experiment, the
cat to be in this state is both alive and died simultaneously. In our perception it
appears in a definite state due to an unavoidable localization process (see \cite{Gh1}
and references therein) to "macro-objectificate" the Cat state. This process is {\it
irreversible}. Its rate is fast for a macro-system, but very slow for a single
electron. So that quantum dynamics of a single electron is not disturbed by this
process.

However, it is legitimate to ask: "How to prepare the Cat state which is needed in a
macro-objectification?" The point is that the cat is evident to be definitely alive
before preparing the state (\ref{1}). In fact its preparation is a
"macro-disobjectification" process to be opposite to the macro-objectification one.

We could imagine this process as follows. At the first stage the electron's source (to
emit only one electron, in each experiment), the vial with a poison and the cat are in
a closed box, but the source of electrons has not yet been switched on. Thus, the cat
is definitely alive at this stage. At the second stage, an observer to be outside of
the box launches the electron's source. As a result, the state of an electron emitted
begins evolving from the initial one to the superposition
$\psi_{tr}^{end}(x,t)+\psi_{ref}^{end}(x,t)$ to describe the electron after the
scattering event. (Remind that, saying about the OI's Cat state, we should adhere to
the OI's attitude toward the wave function).

We might expect that namely at this stage the cat becomes neither alive nor died.
However, the question of how the scattered OI's electron "macro-disobjectificates"
(delocalizes) the cat's state $|1\rangle_c$ arises. Indeed, any direct influence of
the electron (micro-object) on the cat (macro-object) is negligible, hence a
"macro-disobjectification process" implies the existence in Nature of some
communicator (amplifier) to transfer the 'nonlocality' and 'observer-dependence' of
the scattered OI's electron to the yet alive cat. However, such "transfer" looks quite
unrealistic. On the road from the micro- to macro-level of the OI's universe, all
unavoidable stochastic physical processes could facilitate anything but a
"macro-disobjectification process".

So that "the OI's Cat state", i.e., the state where the cat is both alive and died
simultaneously, cannot be prepared in principle: the current interpretation of Cat
states has simply no physical sense. This fact shows that the conflict between the
"OI's Cat states" and the macro-world is much deeper than it has been considered
before. As is seen, not only the measurement problem but also the preparation one
arises for such states. Moreover, while the {\it measurement} problem for the "OI's
Cat states" could be resolved by means of introducing some macro-objectification
(localization) process, the {\it preparation} one is unsolvable in principle.

So that at present the Cat paradox has no consistent solution. To keep the universe as
a knowable real physical world governed both by the superposition principle and the
PMRs, we must solve it at the micro-level. The Cat state (\ref{1}) to imply a {\it
causal} relationship between an electron and cat is needed in extending the
application of the PMRs to a single electron. In this case, the cat is {\it either}
alive {\it or} died because a scattered electron is {\it either} reflected {\it or}
transmitted by a potential barrier.

At first sight, we have arrived at {\it a priori} deadlock conclusion. The point is
that the PMRs are known to be incompatible with nonlocality. At the same time, at the
micro-level, nonlocality is now considered as an experimental fact. As was said in
\cite{Gh1}, "\ldots one must recognize that natural phenomena exhibit basic nonlocal
features, this conclusion being completely independent from the formulation and/or the
interpretation of the theory and stemming simply from the experimental predictions of
QM\ldots"

Nevertheless, despite this dictum, this is not the case. The EPR-Bell experiments -
the main witnesses of nonlocality - do not at all discard Bell's assumption on the
existence of local hidden variables. They discard another (implicit) assumption to
underlie Bell's theorem, QM and EPR-Bell experiments themselves.

\subsection{What do the EPR-Bell experiments falsify, in reality?} \label{Bell}

One has to take into account that all statistical experiments, including EPR-Bell
ones, consist at least of three stages: 1) obtaining experimental data; 2) their
sampling; 3) their averaging and subsequent interpretation.

Of course, the first stage of the EPR-Bell experiments is beyond doubts. However,
already the second one raises questions. Indeed, as it has been shown in \cite{Kh1},
raw data obtained in the optical EPR-Bell experiments \cite{Wei}, examined under the
fair sampling assumption, impugn the "non-signaling principle". By \cite{Kh1}, either
the fair sampling assumption or the "non-signaling principle" is wrong.

However, we have to stress that the fair sampling assumption is in fact a {\it
requirement} for statistical experiments, while the "non-signaling principle", in the
case of the EPR-Bell ones, is only a {\it theoretical prediction} whose validity is
needed in verification. Thus, in fact the study \cite{Kh1} shows that "nonlocal
correlations" to appear in the EPR-Bell experiments \cite{Wei} do not obey the
"non-signaling principle" and, thus, these experiments falsify the theorems to support
nonlocality.

Of course, one could doubt the analysis \cite{Kh1}, because the theorems are certainly
logically consistent and based strictly on the postulates of QM. However, on the other
hand, the study \cite{Kh1} is based on reliable experimental results sampled fairly.
Hence, there is no reason to doubt it. By our approach, namely the theorems to support
the "non-signaling principle" are erroneous because the quantum-mechanical postulates
to underlie theirs are inapplicable, in the current formulation, to Cat states:
quantum theory of Cat states must be based on the PMRs.

Note that the violation of the "non-signaling principle" in the experiments \cite{Wei}
does not at all mean that they indeed dealt with the faster than light signals. All
EPR-Bell experiments do not imply direct measurements of the signal's velocity. They
are aimed only at checking the validity of Bell's inequality, and namely the fact of
its violation is interpreted as a falsity of Bell's assumption on the existence of
local hidden variables.

The crucial stage of all such experiments is just that of averaging the experimental
data. Namely this stage of the EPR-Bell experiments is a loophole for 'nonlocality'
and 'observer-dependence'. Based (like \cite{Gh2}) on the implicit assumption that the
current quantum-mechanical practice of treating Cat states is valid, these experiments
resort to the averaging over the whole two-electron Cat state. However, according to
the PMRs, such averaging has no physical sense. All observables can be introduced only
for macroscopically distinct (alternative) sub-states to enter the Cat state.

Strictly speaking, the EPR-Bell experiments test not only the validity of Bell's
assumption on the existence of local hidden variables, but also the validity of the
{\it implicit} assumption that the current quantum-mechanical practice of averaging
over Cat states is legitimate. Thus, holding in respect 'locality' and
'observer-independence' as inherent properties of the universe, we conclude that {\it
the EPR-Bell experiments based on the fair sampling assumption falsify the current
practice of introducing observables for Cat states, rather than Bell's assumption on
the existence of local hidden variables}.

We have to note that our criticism of the current practice of application of Born's
averaging rule to Cat states complements and develops the approaches (see
\cite{Hes,Kh2} and references therein; see also \cite{Kr1}) which are focused on the
analysis of Bell's theorem itself. They point to Vorob'ev's theorem \cite{Vo1} in
probability theory, which forbids averaging over hidden variables ascribed to
different (nonidentical) sets of EPR-Bell experiments with differently oriented
detectors. The relevance of this theorem to the problem of nonlocality has been,
perhaps, first recognized by W. Philipp (see \cite{Hes,Kh2}).

So, to introduce the PMRs into QM is a reasonable way out in solving the Cat paradox.
Our next step is to show that QM is quite compatible with the PMRs. However, the
natural background for them is the statistical interpretation of QM (and its attitude
toward the one-particle wave function), rather than the OI.

\subsection{The Cat paradox from the viewpoint of the statistical interpretations of QM}
\label{stat}

Note, the OI's attitude toward the one-particle wave function is inconsistent
basically: the wave function cannot be in principle associated with a particle in a
single run, because the former evolves deterministically while the latter behaves
stochastically.

In making choice of a true interpretation of QM, we have to take into account that the
destination of any physical theory is to predict and explain the behavior of the
element of reality, which is under its study. In this context, only the ensemble's, or
statistical interpretation (SI) of QM (e.g., see \cite{Ba2}) reflects properly the
nature of the wave function, as what to describe the state of the {\it ensemble} of
identically prepared particles, i.e., a particle in the infinite set of identical
independent one-particle experiments (unlike ensembles of many-particles systems, a
one-particle quantum ensemble is known can also be realized as a rare {\it beam} of
identically prepared particles to move independently on each other). {\it Elements of
reality, being under study in QM, are quantum ensembles: all its predictions and
rules, as well as their experimental verification, are related just to quantum
ensembles, which behave deterministically}.

By the SI an electron, at the scales much larger than its classical radius, is a
point-like object to move stochastically. It must obey the PMRs and hence {\it the Cat
paradox, within the SI, is not a macro-objectification problem}. On the contrary,
quantum ensembles of electrons are wave-like objects to evolve deterministically. QM
does not explain the appearance of wave properties of the electron's ensembles. One
may only suppose that they are rooted in the structure of a single electron. However,
such an explanation (which would be of great importance) is, perhaps, the prerogative
of a more general, sub-quantum theory.

By the SI, the squared modulus of the wave function, in the $x$- and
$p$-representations, gives the $x$- and $p$-distributions of electrons in a quantum
ensemble. These distributions are connected to each other, because the wave functions
in these representations are the Fourier-transforms of each other. Their connection is
also reflected in the uncertainty relations to tell us that the smaller the size of
the electron's ensemble in the $x$-space, the larger it in the $p$-space; and vise
versa. Here it is relevant to stress once more that the uncertainty relations, like
any other verifiable rule of QM, do not relate to a single electron in a single
experiment!

The SI implies that, in a single one-particle experiment, an observer can measure
(directly or no) a {\it true} random value of the particle's position (or momentum).
The infinite set of identical independent measurements of the particle's position (or
momentum) gives the $x$- (or $p$-)distribution for the ensemble under study, what
allows an observer to verify the $x$- (or $p$-)distribution obtained from the wave
function. So that, {\it within the SI, the Cat paradox is not a measurement problem}.

\section{The Cat paradox as a correspondence problem. Concepts of combined and
elementary quantum processes and states} \label{stat1}

Our analysis of the state $\psi_{tr}^{end}(x,t)+\psi_{ref}^{end}(x,t)$, on the basis
of the SI, suggests that this paradox should be treated as a {\it correspondence}
problem. Indeed, a point-like electron cannot occupy two or more macroscopically
distinct spatial regions, both in classical and quantum mechanics. So that putting a
ban on the "either-or" scenario, in the current interpretation of this (micro-)Cat
state, is simply inconsistent with the CP.

It is relevant also to stress that, for this state, QM as it stands gives not only a
false interpretation, but also {\it a priori} false predictions. Indeed, by the
current understanding of the CP, {\it any} one-particle time-dependent wave function
is a counterpart to a single classical one-particle trajectory. Hence, averaging over
{\it any} one-particle wave function should give the {\it expectation} (i.e., the most
probable) values of one-particle observables. This correspondence rule should be valid
for the superposition $\psi_{tr}^{end}(x,t)+\psi_{ref}^{end}(x,t)$, too. However, it
is easy to show that averaging the electron's position and momentum over this state
does not give the {\it expectation} values of these one-particle observables. So that
the above correspondence rule must be revised.

Namely, QM must distinguish a pure one-particle (sub-)ensemble which has available to
it two or more macroscopically distinct states but occupies at any given time a
definite one of those states (here we rephrase one of Leggett's principles
\cite{Le1}). Such ensembles, their states and corresponding processes will be here
referred to as elementary. Otherwise, we deal with {\it combined} ones.

By the CP, only an {\it elementary} time-dependent one-particle pure state can be
considered as a quantum counterpart to a single one-particle trajectory in classical
mechanics. As regards a {\it combined} time-dependent one-particle pure state, it
represents a coherent superposition of $N$ ($N>1$) macroscopically distinct
(alternative) {\it elementary} sub-states. It is a counterpart to $N$ classical
one-particle trajectories.

The CP forbids application of Born's averaging rule to combined states. One can
introduce one-particle observables only for {\it elementary} ones. Ignoring this
prohibition leads inevitably to paradoxes. All this implies the availability of
experimental schemes to allow measuring observables for each elementary sub-state,
without destroying interference between them.

Thus, in QM based on the CP, Cat states are combined ones and a 1D completed
scattering is a combined one-particle process to consist from two elementary
one-particle sub-processes - transmission and reflection. In the last case all
one-particle observables, including characteristic times, can be introduced only for
either sub-process, separately.

At this point it is relevant to stress that quantum description of combined states,
based on the CP, must provide (\i) rules of decomposing such states into elementary
sub-states and also (\i\i) experimental schemes to allow measuring observables of
elementary sub-states, without destroying the interference pattern. However, at
present QM does not obey this requirement. For example, the standard model of a 1D
completed scattering does not suggest decomposing this process into sub-processes.
Moreover, it is a commonplace that the linear formalism of QM does not allow, in
principle, such a decomposition. In this connection, of importance is a macrorealistic
model \cite{Ch1,Ch2} of this process, which shows that this is not the case.

\section{A macrorealistic model of a 1D completed scattering} \label{model}
\subsection{Wave functions for transmission and reflection}

Note that the model \cite{Ch1,Ch2} deals with an electron to impinge, from the left, a
symmetric potential barrier localized in the spatial region $[a,b]$.

Let $\Psi_{full}(x;E)$ be the wave function to describe the whole ensemble of
identical electrons with energy $E$ ($E=\hbar^2 k^2/(2m)$); to the left of the barrier
-
\[\Psi_{full}(x;E)=\exp(ikx)+A_{full}^{R}\exp(-ikx);\] to the right of the barrier -
\[\Psi_{full}(x;E)=A_{full}^{T}exp(ikx);\] here $A_{full}^{R}$ and $A_{full}^{T}$ are
the known complex amplitudes of the reflected and transmitted waves, respectively; $x$
is the particle's coordinate.

As is shown in \cite{Ch1}, $\Psi_{full}(x;E)$ can be uniquely presented in the form
\begin{eqnarray} \label{3}
\Psi_{full}(x;E)=\Psi_{tr}(x;E)+\Psi_{ref}(x;E);
\end{eqnarray}
$\Psi_{tr}(x;E)$ and $\Psi_{ref}(x;E)$ are solutions of the Schr\"odinger equation to
obey the boundary conditions (\ref{5}). To the left of the barrier, we have
\begin{eqnarray} \label{4}
\Psi_{tr}(x;E)=A_{tr}^{In}\exp(ikx)+A_{tr}^{R}\exp(-ikx),\nonumber\\
\Psi_{ref}(x;E)=A_{ref}^{In}\exp(ikx)+A_{ref}^{R}\exp(-ikx);
\end{eqnarray}
\begin{eqnarray} \label{5}
A_{tr}^{R}=0,\ooa A_{ref}^{R}=A_{full}^{R},\ooa A_{tr}^{In}+A_{ref}^{In}=1,\ooa
|A_{tr}^{In}|=|A_{full}^{T}|,\ooa |A_{ref}^{In}|=|A_{full}^{R}|.
\end{eqnarray}

Note, there are two sets of the amplitudes $A_{tr}^{In}$ and $A_{ref}^{In}$ to satisfy
the boundary conditions (\ref{5}). One of them leads to the wave function
$\Psi_{ref}(x;E)$ to be even, with respect to the point $x_c$ ($x_c=(a+b)/2$). Another
leads to an odd function. We choose the latter. In this case, $\Psi_{ref}(x_c;E)=0$
for any value of $E$. And, at any value of $t$, wave packets formed from the odd
solutions are equal to zero at this point, too. This means that electrons to impinge a
symmetric potential barrier, from the left, do not enter the spatial region $x>x_c$.

Note that both functions, $\Psi_{tr}(x;E)$ and $\Psi_{ref}(x;E)$, contain the terms to
describe electrons impinging the barrier from the right, which are cancelled in the
superposition (\ref{3}). As a result, in this superposition, electrons impinging the
barrier from the left and then being reflected (transmitted) by its are described by
the function $\psi_{ref}(x;E)$ ($\psi_{tr}(x;E)$) where
\begin{eqnarray*}
\psi_{ref}(x;E)\equiv \Psi_{ref}(x;E),\ooo \psi_{tr}(x;E)\equiv \Psi_{tr}(x;E),\ooo
x\leq x_c;\nonumber\\
\psi_{ref}(x;E)\equiv 0,\ooo \psi_{tr}(x;E)\equiv \Psi_{full}(x;E), \ooo x>x_c.
\end{eqnarray*}
It is evident that $\Psi_{full}(x;E)=\psi_{tr}(x;E)+\psi_{ref}(x;E)$.

Note, the first derivatives of $\psi_{tr}(x;E)$ and $\psi_{ref}(x;E)$ with respect to
$x$ are discontinuous at the point $x_c$. However, the probability current density for
either function is constant everywhere! So that the sum of these functions obeys the
Schr\"odinger equation, but either function obeys the continuity equation. The same
holds for all wave packets formed from these functions.

Let $\Psi_{full}(x,t)$ be a solution of the time-dependent Schr\"odinger equation for
a given initial condition. Let also $\Psi_{tr}(x,t)$ and $\Psi_{ref}(x,t)$ be the
corresponding solutions formed from $\Psi_{tr}(x;E)$ and $\Psi_{ref}(x;E)$,
respectively. Besides, let $\psi_{tr}(x,t)$ and $\psi_{ref}(x,t)$ be the corresponding
wave packets formed from $\psi_{tr}(x;E)$ and $\psi_{ref}(x;E)$. Then we have
\begin{eqnarray*}
\Psi_{full}(x,t)=\Psi_{tr}(x,t)+\Psi_{ref}(x,t)\equiv \psi_{tr}(x,t)+\psi_{ref}(x,t)
\end{eqnarray*}

Namely $\psi_{tr}(x,t)$ and $\psi_{ref}(x,t)$ describe, at all stages of scattering,
the motion of the (to-be-)transmitted and (to-be-)reflected subensembles. Either
function obeys the continuity equation, but their sum obeys the Schr\"odinger one.
Hence the superposition $\psi_{tr}(x,t)+\psi_{ref}(x,t)$, unlike
$\Psi_{tr}(x,t)+\Psi_{ref}(x,t)$, consists from probability waves to {\it interact}
with each other (their interaction disappears in the limit $t\to \infty$).

Note that $\Re\langle\psi_{tr}(x,t)|\psi_{ref}(x,t)\rangle=0$ for any value of $t$.
Therefore, despite the interference between $\psi_{tr}$ and $\psi_{ref}$, we have
\begin{eqnarray} \label{7}
\langle\Psi_{full}(x,t)|\Psi_{full}(x,t)\rangle =\textbf{T}+\textbf{R}=1;
\end{eqnarray}
\[\textbf{T}=\langle\psi_{tr}(x,t)|\psi_{tr}(x,t)\rangle,\ooa
\textbf{R}=\langle\psi_{ref}(x,t)|\psi_{ref}(x,t)\rangle;\] $\textbf{T}$ and
$\textbf{R}$ are constants to be the transmission and reflection probabilities,
respectively.

\subsection{Measurable characteristic times for transmission and reflection}

So, a 1D completed scattering is a {\it combined} process to consist from two
alternative coherently evolved {\it elementary} sub-processes, transmission and
reflection. In this case, to observe the time evolution of the wave packet
$\Psi_{full}(x,t)$ means, in fact, to observe that of the interference pattern formed
by these sub-processes.

However, the main peculiarity of a 1D completed scattering, as a {\it combined}
quantum process, is that it also implies performing experiments for testing the {\it
individual} properties of its sub-processes. In \cite{Ch2}, both for transmission and
reflection, we have defined the time spent, on the average, by an electron in the
barrier region. They are the Larmor times $\tau^L_{tr}$ and $\tau^L_{ref}$ (see
\cite{Ch2}):
\begin{eqnarray} \label{8}
\tau^L_{tr}=\frac{1}{\textbf{T}}\int_{-\infty}^\infty dt\int_a^b dx |\psi_{tr}(x,t)|^2
\equiv \frac{1}{\textbf{T}}\int_0^\infty G(k)T(k)\tau^{dwell}_{tr}(k)dk\\
\tau^L_{ref}=\frac{1}{\textbf{R}}\int_{-\infty}^\infty dt\int_a^{x_c}dx
|\psi_{ref}(x,t)|^2 \equiv \frac{1}{\textbf{R}}\int_0^\infty
G(k)R(k)\tau^{dwell}_{ref}(k)dk \nonumber
\end{eqnarray}
where $G(k)=g(k)-g(-k)$; $g(k)$ is the Fourier-transform of $\Psi_{full}(x,0)$; $T(k)$
and $R(k)$ are the real transmission and reflection coefficients, respectively. The
dwell times for transmission, $\tau^{dwell}_{tr}(k)$, and reflection,
$\tau^{dwell}_{ref}(k)$, are defined by the expressions
\[\tau^{dwell}_{tr}(k)=\frac{m}{\hbar k T(k)}\int_a^b |\psi_{tr}(x,k)|^2dx,\ooo
\tau^{dwell}_{ref}(k)=\frac{m}{\hbar k R(k)}\int_a^{x_c} |\psi_{ref}(x,k)|^2dx.\]

It is crucial that the characteristic times $\tau^L_{tr}$ and $\tau^L_{ref}$ can be
measured with the help of the Larmor clock procedure. This procedure (see \cite{Ch2}
and also \cite{But}) implies switching on an infinitesimal magnetic field in the
barrier region. Then the angle of the Larmor precession of the average electron's spin
is measured separately for the transmitted and reflected subensembles, well after the
scattering event. That is, in this procedure the average electron's spin serves as a
clock-pointer to "remember" the time spent by an electron in the barrier region. It is
evident that all measurements performed well after the scattering event do not distort
interference between these sub-process.

Of importance is also to stress that this procedure does not allow measuring the {\it
asymptotic group times} for transmission and reflection, which differ from the
corresponding Larmor times (see \cite{Ch2}). This fact supports our belief that the
passage of an electron through the barrier region is an observer-independent process
and hence QM must give a unique definition of the time spent by a particle in this
region, for either sub-process. It says that only the Larmor-time concept gives
relevant characteristic times. As regards the group times for these sub-processes,
they cannot be measured in principle, and hence they have no physical sense. This
concerns all other characteristic times deduced from tracing particular points of a
scattered wave packet: in the barrier region, such tracing cannot be performed in
principle, because there is no causal relationship between the incident and
transmitted (or reflected) wave packets.

Note, the Larmor clock procedure allows one to discriminate between our and standard
models of a 1D completed scattering. The latter defines characteristic times on the
basis of the wave function to describe the whole process (see, e.g., \cite{Mu0,But}).
This step violates the CP and, as a result, this model predicts the so called Hartman
effect (superluminal tunneling) whose interpretation (see, e.g., \cite{Win} and
references therein) is extremely moot and its experimental verification is extremely
unreliable (see \cite{Ran}). On the contrary, our model respects the CP and gives a
physically meaningful explanation of the tunneling phenomenon, in a complete agreement
with special relativity.

\section{Conclusion}

On the basis of our macrorealistic model of a 1D completed scattering, we have shown
that the existing disparity between everyday physical reality and the reality of
quantum theory is not real. It results from the fact that QM as it stands violates the
CP. That is, the Cat paradox is a correspondence problem, rather than the measurement
or macro-objectification one. QM {\it must and can} be presented as a macrorealistic
theory to respect the PMRs and hence the CP. In this theory, a CSMDS must be treated
as a {\it combined} one to consist from coherently evolved, macroscopically distinct
elementary sub-states.

Combined states are governed both by the superposition principle and the PMRs, without
any conflict between them. No observables can be introduced for combined states.
Born's averaging rule is applicable only to elementary sub-states. We have to stress
that experimental observations of interference between (sub-)states to constitute a
CSMDS do not at all evidence against the macro-realism of QM. Apart from such
experiments, combined states imply also experiments to allow inspecting the individual
properties of the sub-states, without destroying the interference pattern.

Cat states are combined ones. Thus, there is no room for the Cat paradox in a
macrorealistic QM. As regards the EPR-Bell experiments, they discard the current
practice of averaging over Cat states, rather than Bell's assumption on the existence
of local hidden variables.

\end{document}